\documentstyle[11pt,epsf]{article}
\newtheorem{theorem}{Theorem}

\newcommand {\dfn} {\stackrel{\Delta} {=}}
\newcommand {\exe} {\stackrel{\cdot} {=}}
\newcommand {\lexe} {\stackrel{\cdot} {\le}}
\newcommand {\gexe} {\stackrel{\cdot} {\ge}}

\newcommand {\reals} {{\rm I\!R}}

\newcommand {\bs} {\mbox{\boldmath $s$}}

\newcommand {\bw} {\mbox{\boldmath $w$}}
\newcommand {\bx} {\mbox{\boldmath $x$}}
\newcommand {\by} {\mbox{\boldmath $y$}}

\newcommand {\bE} {\mbox{\boldmath $E$}}

\newcommand {\bS} {\mbox{\boldmath $S$}}

\newcommand {\bW} {\mbox{\boldmath $W$}}
\newcommand {\bX} {\mbox{\boldmath $X$}}
\newcommand {\bY} {\mbox{\boldmath $Y$}}

\newcommand{\calA}{{\cal A}}
\newcommand{\calB}{{\cal B}}

\newcommand{\calE}{{\cal E}}

\newcommand{\calG}{{\cal G}}

\newcommand{\hH}{\hat{H}}
\newcommand{\hP}{\hat{P}}
\newcommand{\hI}{\hat{I}}
\newcommand{\calI}{{\cal I}}

\newcommand{\calS}{{\cal S}}
\newcommand{\calT}{{\cal T}}

\newcommand{\calW}{{\cal W}}
\newcommand{\calX}{{\cal X}}
\newcommand{\calY}{{\cal Y}}

\begin{document}
\thispagestyle{empty}
\title{Ensemble Performance of Biometric Authentication\\ Systems Based on Secret
Key Generation}
\author{Neri Merhav}
\date{}
\maketitle

\begin{center}
The Andrew \& Erna Viterbi Faculty of Electrical Engineering\\
Technion - Israel Institute of Technology \\
Technion City, Haifa 32000, ISRAEL \\
E--mail: {\tt merhav@ee.technion.ac.il}\\
\end{center}
\vspace{1.5\baselineskip}
\setlength{\baselineskip}{1.5\baselineskip}

\begin{abstract}
We study the ensemble performance of biometric authentication systems, 
based on secret key generation, which work as follows. In the enrollment
stage, an individual provides a biometric signal that is mapped into a secret
key and a helper message, the former being prepared to become
available to the system at a later time (for authentication), 
and the latter is stored in a public database. When an authorized user
requests authentication, claiming his/her identity as one of the subscribers, 
s/he has to provide a biometric signal again, and then the system, which
retrieves also the helper message of the claimed subscriber, produces an estimate of the secret
key, that is finally compared to the secret key of the claimed user. In case
of a match, the authentication request is approved, otherwise, it is rejected.
Referring to an ensemble of systems based on Slepian--Wolf binning, we provide
a detailed analysis of the false--reject and false--accept probabilities, for a wide 
class of stochastic decoders. We also comment on the security for the typical
code in the ensemble.\\

\end{abstract}

\section*{I. Introduction}

We consider a biometric authentication system 
that is described in
\cite[Sections 2.2--2.6]{IW10}, which is based on the notion of secret key
generation and sharing due to Maurer \cite{Maurer93}
and Ahlswede and Csisz\'ar \cite{AC93}, \cite{AC98}. Specifically,
such a system works as follows. In the enrollment
stage, an individual which subscribes to the system provides
a biometric signal, $\bX =(X_1,X_2,\ldots,X_n)$. The system receives this
signal and generates (using its encoder) two outputs in response. The first output is a secret
key, $\bS$, at rate $R_{\mbox{\tiny s}}$ 
and the second is a helper message, $\bW$, 
at rate $R_{\mbox{\tiny w}}$. The secret key is prepared in
order to be used 
by the system later, at the authentication stage. The helper message
is stored in a public database. When an authorized user (a subscriber)
wishes to sign in, claiming his/her identity as one of the existing subscribers,
s/he is requested to provide again his/her biometric signal,
$\bY=(Y_1,\ldots,Y_n)$ (correlated to $\bX$, if indeed from the same individual, or
independent, if not). The system then
retrieves the helper message $\bW$ of the claimed subscriber, 
and responds (using its decoder) by estimating the secret
key, $\hat{\bS}$ (based on $(\bY,\bW)$), and comparing it to the secret key of the
claimed user, $\bS$. In case
of a match, access to the system is granted, otherwise, it is denied. 

In \cite[Sect.\ 2.3]{IW10}, achievable rate pairs $(R_{\mbox{\tiny s}},
R_{\mbox{\tiny w}})$ were found for the existence of systems (encoders and
decoders) that satisfy the
following three requirements in the large $n$ limit: (i) arbitrarily small false--reject (FR)
probability, (ii) arbitrarily small false--accept (FA) probability, and
(iii) arbitrarily small leakage between the secret message and the helper
message, in terms of the asymptotic normalized mutual information, 
$I(\bS;\bW)/n$. In particular, Theorem 2.1 of
\cite{IW10} asserts that when $(\bX,\bY)$ are drawn from a discrete memoryless
source (DMS), generating independent copies of a correlated pair $(X,Y)\sim
P_{XY}$, the maximum achievable key rate, $R_{\mbox{\tiny s}}$, 
under the above constraints,
is given by the single--letter mutual information, $I(X;Y)$. 
It then follows that $R_{\mbox{\tiny w}}$ must
lie in the range $H(X|Y) < R_{\mbox{\tiny
w}} < H(X)-R_{\mbox{\tiny s}}$, 
where the conditional entropy in the lower limit is essential for
reliable identification of an authorized subscriber 
(small FR probability) and the upper limit is
essential for the security requirement. These limitations already guarantee
that $R_{\mbox{\tiny w}} < H(X)$, which is essential for keeping
the FA probability vanishingly small for large $n$.

As in many proofs of direct coding theorems in the information theory literature, 
in the achievability part of 
\cite[Theorem 2.1]{IW10} too, the analyses of the error probabilities
(in this case, the FA and the FR probabilities) are very rough -- they are
merely good enough to prove the achievability of the desired coding rates
in the simplest possible manner. However, these are poor estimates of the
achievable FR and FA probabilities themselves when these are considered to be the
relevant performance metrics for given $R_{\mbox{\tiny s}}$ and
$R_{\mbox{\tiny w}}$.

The purpose of this paper is to provide sharper evaluations of the ensemble
performance of the FA and the FR probabilities. 
In particular, referring to an ensemble 
of systems based on Slepian--Wolf binning, we provide
detailed analyses of the exponential behavior of the FR probability, for
a wide class of stochastic decoders, which includes the respective maximum
a posteriori (MAP) decoder as a special case. An expurgated bound is provided as
well and discussed quite in detail. For the FA probability, we analyze the
ensemble performance of the MAP decoder and provide some intuition concerning
its behavior.
We also comment on the security of the code for the typical
code in the ensemble.

The paper is organized as follows. 
In Section II, we establish the notation conventions.
In Section III, we formalize the setup and
spell out the objectives. In Section IV, we present and discuss the random
coding FR exponent 
and an expurgated bound. In Section V, we derive the
random coding FA exponent, and finally, in Section VI, we discuss the leakage
of the typical code.

\section*{II. Notation Conventions}

Throughout the paper, random variables will be denoted by capital
letters, specific values they may take will be denoted by the
corresponding lower case letters, and their alphabets
will be denoted by calligraphic letters. Random
vectors and their realizations will be denoted,
respectively, by capital letters and the corresponding lower case letters,
both in the bold face font. Their alphabets will be superscripted by their
dimensions. For example, the random vector $\bX=(X_1,\ldots,X_n)$, ($n$ --
positive integer) may take a specific vector value $\bx=(x_1,\ldots,x_n)$
in $\calX^n$, the $n$--th order Cartesian power of $\calX$, which is
the alphabet of each component of this vector.
Sources and channels will be denoted by the letter $P$ or $Q$,
subscripted by the names of the relevant random variables/vectors and their
conditionings, if applicable, following the standard notation conventions,
e.g., $Q_X$, $P_{Y|X}$, and so on. When there is no room for ambiguity, these
subscripts will be omitted.
The probability of an event $\calG$ will be denoted by $\mbox{Pr}\{\calG\}$,
and the expectation
operator with respect to (w.r.t.) a probability distribution $P$ will be
denoted by
$\bE_P\{\cdot\}$. Again, the subscript will be omitted if the underlying
probability distribution is clear from the context.
The entropy of a generic distribution $Q$ on $\calX$ will be denoted by
$H_Q(X)$. For two
positive sequences $a_n$ and $b_n$, the notation $a_n\exe b_n$ will
stand for equality in the exponential scale, that is,
$\lim_{n\to\infty}\frac{1}{n}\log \frac{a_n}{b_n}=0$. Similarly,
$a_n\lexe b_n$ means that
$\limsup_{n\to\infty}\frac{1}{n}\log \frac{a_n}{b_n}\le 0$, and so on.
The indicator function
of an event $\calG$ will be denoted by $\calI\{\calG\}$. The notation $[x]_+$
will stand for $\max\{0,x\}$.

The empirical distribution of a sequence $\bx\in\calX^n$, which will be
denoted by $\hat{P}_{\bx}$, is the vector of relative frequencies
$\hat{P}_{\bx}(x)$
of each symbol $x\in\calX$ in $\bx$.
The type class of $\bx\in\calX^n$, denoted $\calT(\hP_{\bx})$, is the set of all
vectors $\bx'$
with $\hat{P}_{\bx'}=\hat{P}_{\bx}$. 
Information measures associated with empirical distributions
will be denoted with `hats' and will be subscripted by the sequences from
which they are induced. For example, the entropy associated with
$\hat{P}_{\bx}$, which is the empirical entropy of $\bx$, will be denoted by
$\hat{H}_{\bx}(X)$. 
Similar conventions will apply to the joint empirical
distribution, the joint type class, the conditional empirical distributions
and the conditional type classes associated with pairs (and multiples) of
sequences of length $n$.
Accordingly, $\hP_{\bx\by}$ will be the joint empirical
distribution of $(\bx,\by)=\{(x_i,y_i)\}_{i=1}^n$,
and $\calT(\hP_{\bx\by})$ will denote
the joint type class of $(\bx,\by)$. Similarly, $\calT(\hP_{\bx|\by}|\by)$ 
will stand for
the conditional type class of $\bx$ given
$\by$, $\hH_{\bx\by}(X,Y)$ 
will designate the empirical joint entropy of $\bx$
and $\by$,
$\hH_{\bx\by}(X|Y)$ will be the empirical conditional entropy,
$\hI_{\bx\by}(X;Y)$ will
denote empirical mutual information, and so on.
We will also use similar rules of notation in the context of
a generic distribution, $Q_{XY}$ (or $Q$, for short): we use
$\calT(Q_X)$ for the type class of sequences with empirical distribution $Q_X$,
$H_Q(X)$ -- for the corresponding empirical entropy, 
$\calT(Q_{XY})$ -- for the joint type class x,
$T(Q_{X|Y}|\by)$ -- for the conditional type class of $\bx$ given $\by$,
$H_Q(X,Y)$ -- for the joint empirical entropy, 
$H_Q(X|Y)$ -- for the conditional empirical entropy, 
$I_Q(X;Y)$ -- for the empirical mutual information, and so on.
We will also use the customary notation for the weighted divergence,
\begin{equation}
D(Q_{Y|X}\|P_{Y|X}|Q_X)=\sum_{x\in\calX}Q_X(x)\sum_{y\in\calY}Q_{Y|X}(y|x)\log
\frac{Q_{Y|X}(y|x)}{P_{Y|X}(y|x)}.
\end{equation}

\section*{III. Setup and Objectives}

Consider the following system model 
for biometric identification. An {\it enrollment source} sequence,
$\bx=(x_1,\ldots,x_n)$, which is a realization of the random vector
$\bX=(X_1,\ldots,X_n)$,
that emerges from a discrete memoryless source (DMS),
$P_X$, with a finite alphabet $\calX$, is fed into an {\it enrollment
encoder}, $\calE$, that produces two
outputs: a secret key, $\bs$ (a realization of a random variable
$\bS$), and a helper 
message, $\bw$ (a realization of $\bW$), taking on values in finite alphabets,
$\calS_n=\{0,1,\ldots,e^{nR_{\mbox{\tiny s}}}\}$ and
$\calW_n=\{0,1,\ldots,e^{nR_{\mbox{\tiny w}}}\}$, respectively, where
$R_{\mbox{\tiny s}}$ is the {\it secret--key rate}, and
$R_{\mbox{\tiny w}}$ is the {\it helper--message rate}.
This encoding operation designates the enrollment stage.

We consider the ensemble of enrollment encoders, 
$\{\calE\}$, generated by {\it random binning}, where
for each source vector $\bx\in\calX$, one selects independently at random,
both a secret key and a helper message, under the uniform distributions across
$\calS_n$ and $\calW_n$, respectively. In other words, denoting by $\bw=f(\bx)$ and
$\bw=f(\bx)$, the randomly selected bin assignments for both outputs, it is
assumed that the $2|\calX|^n$ random variables
$\{f(\bx),g(\bx)\}_{\bx\in\calX^n}$ are all mutually independent.

The {\it authentication decoder}, $\calA$, which is aware of the randomly
selected encoder, $\calE$, is fed by two inputs: the helper message $\bw$ and an {\it
authentication source} sequence, $\by=(y_1,\ldots,y_n)$ (a realization of
$\bY=(Y_1,\ldots,Y_n)$), that is produced at 
the output of a discrete memoryless channel (DMC),
$P_{Y|X}$, with a finite output alphabet $\calY$, that is fed by $\bx$.
The output of the authentication decoder 
is $\hat{\bs}=U(\by,\bw)$ (a realization of $\hat{\bS}$), which is an estimate 
(possibly, randomized) of the
secret key, $\bs$. If $\hat{\bs}=\bs$, access to the system is granted,
otherwise, it is denied. This decoding operation stands for the authentication
stage.

The optimal estimator of $\bs$, based on $(\by,\bw)$, in the sense of
minimum FR probability, $\mbox{Pr}\{\hat{S}\ne\bS\}$, is 
the maximum a posteriori probability (MAP) estimator, given by
\begin{equation}
\hat{\bs}_{\mbox{\tiny MAP}}=U(\by,\bw)\dfn\mbox{arg}\max_{\bs} P(\bs,\bw|\by)=
\mbox{arg}\max_{\bs}
\sum_{\bx\in\calX^n}P(\bx|\by)\cdot\calI\{f(\bx)=\bw\}\cdot\calI\{g(\bx)=\bs\},
\end{equation}
where $P(\bx|\by)$ (shorthand notation for $P_{\bX|\bY}(\bx|\by)$) 
is the posterior probability of $\bX=\bx$ given $\bY=\by$,
that is induced by the product distribution, $P_{XY}$ (and the
subscript $XY$ will sometimes be suppressed for simplicity, 
when there is no risk of compromising clarity).

In this paper, we expand the scope and study a more general class of
decoders. This is a class of generalized stochastic likelihood decoders
\cite{p187}, \cite{SMF15}, \cite{SCP14}, \cite{YAG13}, where the decoder
randomly selects its estimate $\hat{\bs}$ according to the posterior
distribution
\begin{equation}
\label{gld}
\tilde{P}(\bs|\by,\bw)=\frac{\sum_{\bx\in\calX^n}
\exp\{na(\hP_{\bx\by})\}\cdot\calI\{f(\bx)=\bw\}\cdot\calI\{g(\bx)=\bs\}}
{\sum_{\bx\in\calX^n}
\exp\{na(\hP_{\bx\by})\}\cdot\calI\{f(\bx)=\bw\}},
\end{equation}
where the function $a(\cdot)$, henceforth referred to as the {\it decoding
metric}, is an
arbitrary continuous function of the joint empirical
distribution $\hP_{\bx\by}$. Throughout the sequel, we will refer to the numerator of the
r.h.s.\ as
$\tilde{P}(\bs,\bw|\by)$, and to the denominator as
$\tilde{P}(\bw|\by)$. For
\begin{equation}
\label{ordinary1}
a(\hP_{\bx\by})=\sum_{x\in\calX}\sum_{y\in\calY}\hP_{\bx\by}(x,y)\ln
P(x|y),
\end{equation}
we have the ordinary likelihood decoder in spirit of \cite{SMF15},
\cite{SCP14}, \cite{YAG13}. For
\begin{equation}
\label{ordinary2}
a(\hP_{\bx\by})=\beta\sum_{x\in\calX}\sum_{y\in\calY}\hP_{\bx\by}(x,y)\ln
P(x|y),
\end{equation}
$\beta$ being a free parameter (sometimes referred to as the inverse
temperature parameter \cite{Rujan93} due to the analogy in statistical
mechanics), we
extend this likelihood decoder to a parametric family of decoders, where
$\beta$ controls the skewedness of the posterior. In particular,
$\beta\to\infty$ leads to the ordinary MAP decoder, $\hat{\bs}_{\mbox{\tiny
MAP}}$. Other interesting choices are associated with mismatched metrics,
\begin{equation}
\label{mismatched}
a(\hP_{\bx\by})=\sum_{x\in\calX}\sum_{y\in\calY}\hP_{\bx\by}(x,y)\ln
P^{\prime}(x|y),
\end{equation}
$P^{\prime}$ being different from $P$, and
\begin{equation}
\label{minimumentropy}
a(\hP_{\bx\by})=-\beta\hH_{\bx\by}(X|Y),
\end{equation}
which for $\beta\to\infty$
approaches the universal minimum entropy decoder (see also discussion around
eqs.\ (5)--(7) of \cite{p187}).

An unauthorized user (i.e., an imposter), 
who claims for a given subscriber identity and wishes to break
into the system, does not
have the correlated biometric data $\by$. The best s/he can do is to estimate $\bs$
based on the only data s/he has, which is the helper message $\bw$, and then
forges any fake biometric data $\tilde{\by}$, 
which together with $\bw$, would cause the
decoder to output this estimate of $\bs$. More precisely, the imposter first
estimates $\bs$ according to
\begin{equation}
\label{imposter}
\tilde{\bs}=V(\bw)\dfn\mbox{arg}\max_{\bs} P(\bs|\bw)=
\mbox{arg}\max_{\bs}
\sum_{\bx\in\calX^n}P(\bx)\cdot\calI\{f(\bx)=\bw\}\cdot\calI\{g(\bx)=\bs\},
\end{equation}
and then generates any $\tilde{\by}\in\calY^n$ 
such that $U(\tilde{\by},\bw)=\tilde{\bs}$, and uses it as the
biometric signal for authentication.

The objectives of the paper are to obtain: (i) ensemble--tight, exponential error
bounds for the average FR probability, $\bar{P}_{\mbox{\tiny
FR}}=\mbox{Pr}\{\hat{\bS}\ne \bS\}$, associated with the generalized stochastic
likelihood decoder (\ref{gld}), as well as an expurgated bound following the
methodology of \cite[Theorem 2]{p187} (see also the correction
\cite{p187correction}), and (ii) an exponential error bound for the average FA
probability of (\ref{imposter}), $\bar{P}_{\mbox{\tiny
FA}}=\mbox{Pr}\{\tilde{\bS}=\bS\}$. Finally, we provide an outline
of a derivation of the leakage, $I(\bS;\bW)$, for a typical code,
$\calE$, in the large $n$ limit.

\section*{IV. False--Reject Error Analysis}

\subsection*{A. Random Coding Exponent}

Consider the system configuration described in Section III, along with the
generalized stochastic likelihood decoder (\ref{gld}).
Define the functions
\begin{equation}
\label{ERQ}
E(R_{\mbox{\tiny w}},Q_{X_0Y})\dfn\min_{Q_{X|Y}}
[R_{\mbox{\tiny w}}-H_Q(X|Y)+
[a(Q_{X_0 Y})-a(Q_{XY})]_+]_+
\end{equation}
and 
\begin{equation}
\label{ErR}
E_{\mbox{\tiny r}}^{\mbox{\tiny FR}}(R_{\mbox{\tiny w}})\dfn
\min_{Q_{X_0Y}}\{D(Q_{X_0Y}\|P_{XY})+E(R_{\mbox{\tiny
w}},Q_{X_0Y})\}.
\end{equation}

Our first result is the following.
\begin{theorem}
\label{fr}
Consider the system configuration described in Section III. Then,
\begin{equation}
\lim_{n\to\infty}\left[-\frac{\ln\bar{P}_{\mbox{\tiny
FR}}}{n}\right]=E_{\mbox{\tiny r}}^{\mbox{\tiny FR}}(R_{\mbox{\tiny w}}).
\end{equation}
\end{theorem}
Before providing the proof, a few points should be discussed.\\

\noindent
1. First, observe that Theorem \ref{fr} asserts that $E_{\mbox{\tiny
r}}^{\mbox{\tiny FR}}(R_{\mbox{\tiny w}})$
is the {\it exact} random coding FR exponent, 
not just a lower bound. This is due to the fact
that all steps of the analytic derivation are ensemble--tight in the
exponential scale, thanks to the ability to avoid the use of the Jensen
inequality and other well known tools that are traditionally 
used to facilitate the analysis, at the
possible price of compromising tightness (see the proof of Theorem \ref{fr}
below).\\

\noindent
2. It is interesting to observe that the FR random coding exponent,
$E_{\mbox{\tiny r}}^{\mbox{\tiny FR}}(R_{\mbox{\tiny w}})$, depends only on
$R_{\mbox{\tiny w}}$, not on
$R_{\mbox{\tiny s}}$. This fact is not trivial, but the intuition is the
following: in
order to estimate $\bS$ correctly, with high probability, from the given data
$(\bY,\bW)$, there should be essentially no ambiguity, first of all, in
defining what the correct $\bS$ is. This will be the case if there is
essentially only one source vector $\bX$ that is responsible for the given
$\bW$ and then this $\bX$ would dictate the correct $\bS=g(\bX)$. 
This in turn would happen
with high probability as long as $R_{\mbox{\tiny w}} > H(X|Y)$.
Otherwise, if more
than one source vector (in the same conditional type class given
$\bY$ as the correct one) is mapped by the encoder to the same helper message, 
then at least one such source vector is likely to be mapped
to a different secret key message, and then the decoding would be ambiguous.
It appears then that correct
estimation of $\bS$ is essentially equivalent to correct estimation of $\bX$,
as in ordinary Slepian--Wolf decoding \cite{Gallager76} (see also
\cite{WM15} and references therein), where there is no
secret key at all (or alternatively, $R_{\mbox{\tiny s}}\to\infty$).
Indeed, the Slepian--Wolf coding component of the joint source--channel coding
system, analyzed in \cite[Section IV]{p187} under the generalized likelihood
decoder, contributes the very same error
exponent as asserted in Theorem \ref{fr}.\\

\noindent
3. It is interesting to examine a few decoding metrics. Consider the choice
$a(Q)=-H_Q(X|Y)$. In this case, we have
\begin{eqnarray}
& &\min_{Q_{X|Y}}[R_{\mbox{\tiny w}}-H_Q(X|Y)+
[a(Q_{X_0 Y})-a(Q_{XY})]_+]_+\nonumber\\
&=&\min_{Q_{X|Y}}[R_{\mbox{\tiny w}}-H_Q(X|Y)+
[H_Q(X|Y)-H_Q(X_0|Y)]_+]_+\nonumber\\
&=&\min_{Q_{X|Y}}[R_{\mbox{\tiny w}}-\min\{H_Q(X|Y),H_Q(X_0|Y)\}]_+\nonumber\\
&=&[R_{\mbox{\tiny w}}-\min\{\max_{Q_{X|Y}}H_Q(X|Y),H_Q(X_0|Y)\}]_+\nonumber\\
&=&[R_{\mbox{\tiny w}}-H_Q(X_0|Y)]_+,
\end{eqnarray}
which, together with (\ref{ErR}), yields the same random coding exponent as the
optimal MAP decoder for Slepian--Wolf decoding (see also
\cite{p187} and \cite{SMF15}). More generally, 
the same comment applies to 
$a(Q)=-\beta H_Q(X|Y)$ for every $\beta\ge 1$, where $\beta\to\infty$
pertains to the deterministic universal minimum entropy decoding, the source--coding dual
to maximum mutual information (MMI) universal decoding (see, e.g., \cite{WM15} and
references therein). For $a(Q)=\beta\bE_Q\ln P(X|Y)$, we have a
finite--temperature likelihood decoder. For $\beta\to\infty$, we are back to
the ordinary MAP decoder, which yields
\begin{eqnarray}
\label{maplimit}
& &\lim_{\beta\to\infty}\min_{Q_{X|Y}}[R_{\mbox{\tiny w}}-H_Q(X|Y)+
[a(Q_{X_0 Y})-a(Q_{XY})]_+]_+\nonumber\\
&=&\lim_{\beta\to\infty}\min_{Q_{X|Y}}[R_{\mbox{\tiny w}}-H_Q(X|Y)+
\beta[\bE_Q\ln P(X_0|Y)-\bE_Q\ln
P(X|Y)]_+]_+\nonumber\\
&=&\min_{\{Q_{X|Y}:~\bE_Q\ln P(X|Y)\ge\bE_Q\ln P(X_0|Y)\}}[R_{\mbox{\tiny
w}}-H_Q(X|Y)]_+,
\end{eqnarray}
which, together with (\ref{ErR}), yields the random coding exponent of the MAP
decoder, as expected. As argued above, this is the same as the exponent achieved by
$a(Q)=-\beta H_Q(X|Y)$ for all $\beta\ge 1$.\\

The remaining part of this section is devoted to the proof of Theorem
\ref{fr}.

\noindent
{\it Proof of Theorem \ref{fr}.}
The expected FR probability is given by
\begin{equation}
\bar{P}_{\mbox{\tiny FR}}=\bE\left\{
\sum_{\bs\ne\bS} \tilde{P}(\bs|\bW,\bY)\right\}
\end{equation}
where the expectation is w.r.t.\ both the randomness of $(\bS,\bW,\bY)$ and
the randomness of the code, $\calE$. For given realizations,
$\bX=\bx$ and $\bY=\by$, let us denote
\begin{equation}
\bar{P}_{\mbox{\tiny FR}}(\bx,\by)\dfn\bE\left\{\sum_{\bs^\prime\ne g(\bx)}
\tilde{P}(\bs^\prime|f(\bx),\by)\right\},
\end{equation}
where now the expectation is merely w.r.t.\ the randomness of $\calE$.
Now, following eq.\ (\ref{gld}),
\begin{eqnarray}
\tilde{P}(\bs^\prime|f(\bx),\by)&=&
\frac{\sum_{\bx^\prime\in\calX^n}
\exp\{na(\hP_{\bx^\prime\by})\}\cdot\calI\{f(\bx^\prime)=f(\bx)\}
\cdot\calI\{g(\bx^\prime)=\bs^\prime\}}
{\sum_{\bx^\prime\in\calX^n}
\exp\{na(\hP_{\bx^\prime\by})\}\cdot\calI\{f(\bx^\prime)=f(\bx)\}}\nonumber\\
&=&\frac{\sum_{Q_{X|Y}}e^{na(Q_{XY})}N(\calT(Q_{X|Y}|\by),f(\bx),\bs^\prime)}
{e^{na(\hP_{\bx\by})}+\sum_{Q_{X|Y}}e^{na(Q_{XY})}N(\calT(Q_{X|Y}|\by),f(\bx))},
\end{eqnarray}
where the summations over $\{Q_{X|Y}\}$ are across all conditional types
$\{\calT(Q_{X|Y}|\by)\}$ of sequences of length $n$, and where
\begin{equation}
N(\calT(Q_{X|Y}|\by),\bw,\bs^\prime)=
\bigg|\calT(Q_{X|Y}|\by)\bigcap
\left\{\bx^\prime:~f(\bx^\prime)=\bw,~g(\bx^\prime)=\bs^\prime
\right\}\bigg|,
\end{equation}
and 
\begin{equation}
N(\calT(Q_{X|Y}|\by),\bw)=
\bigg|\calT(Q_{X|Y}\bigcap\left\{\bx^\prime:~f(\bx)=\bw,~\bx^\prime\ne\bx
\right\}\bigg|.
\end{equation}
Let us first consider the average FR probability for a given $(\bx,\by)$ while
fixing the realizations of $\bw=f(\bx)$ and $\bs=g(\bx)$:
\begin{eqnarray}
\bar{P}_{\mbox{\tiny FR}}(\bx,\by,\bs,\bw)&=&\bE\left\{
\frac{\sum_{\bs^\prime\ne\bs}\sum_{Q_{X|Y}}e^{na(Q_{XY})}
N(\calT(Q_{X|Y}|\by),f(\bx),\bs^\prime)}
{e^{na(\hP_{\bx\by})}+\sum_{Q_{X|Y}}e^{na(Q_{XY})}
N(\calT(Q_{X|Y}|\by),f(\bx))}\right\}\nonumber\\
&=&\int_0^1\mbox{d}t\cdot\mbox{Pr}\left\{\frac{\sum_{Q_{X|Y}}e^{na(Q_{XY})}
N(\calT(Q_{X|Y}|\by),f(\bx))}{e^{na(\hP_{\bx\by})}+
\sum_{Q_{X|Y}}e^{na(Q_{XY})}N(\calT(Q_{X|Y}|\by),f(\bx))}\ge
t\right\}\nonumber\\
&=&n\cdot\int_0^\infty\mbox{d}\theta e^{-n\theta}\cdot
\mbox{Pr}\left\{\frac{\sum_{Q_{X|Y}}e^{na(Q_{XY})}
N(\calT(Q_{X|Y}|\by),f(\bx))}{e^{na(\hP_{\bx\by})}+
\sum_{Q_{X|Y}}e^{na(Q_{XY})}N(\calT(Q_{X|Y}|\by),f(\bx))}\ge
e^{-n\theta}\right\}\nonumber\\
&\exe&\int_0^\infty\mbox{d}\theta e^{-n\theta}\cdot
\mbox{Pr}\left\{\sum_{Q_{X|Y}}e^{na(Q_{XY})}
N(\calT(Q_{X|Y}|\by),f(\bx)) > e^{n[a(\hP_{\bx\by})-\theta]}
\right\}\nonumber\\
&\exe&\int_0^\infty\mbox{d}\theta e^{-n\theta}\cdot
\mbox{Pr}\left\{\max_{Q_{X|Y}}e^{na(Q_{XY})}
N(\calT(Q_{X|Y}|\by),f(\bx)) > e^{n[a(\hP_{\bx\by})-\theta]}
\right\}\nonumber\\
&\exe&\int_0^\infty\mbox{d}\theta e^{-n\theta}\cdot
\mbox{Pr}\bigcup_{Q_{X|Y}}\left\{e^{na(Q_{XY})}
N(\calT(Q_{X|Y}|\by),f(\bx)) > e^{n[a(\hP_{\bx\by})-\theta]}
\right\}\nonumber\\
&\exe&\max_{Q_{X|Y}}\int_0^\infty\mbox{d}\theta e^{-n\theta}\cdot
\mbox{Pr}\left\{
N(\calT(Q_{X|Y}|\by),f(\bx)) > e^{n[a(\hP_{\bx\by})-a(Q_{XY})-\theta]}
\right\}.
\end{eqnarray}
Now, observe that $N(\calT(Q_{X|Y}|\by),f(\bx))$ is a binomial random variable
with $|\calT(Q_{X|Y}|\by)|\exe e^{nH_Q(X|Y)}$ trials and probability of
success $e^{-nR_{\mbox{\tiny w}}}$. Similarly as argued, e.g., in \cite{p187} (see page 5042,
bottom half of the right column therein), we have
\begin{equation}
\mbox{Pr}\left\{
N(\calT(Q_{X|Y}|\by),f(\bx)) > e^{n[a(Q_{X_0Y})-a(Q_{XY})-\theta]}\right\}
\exe e^{-nE(Q_{XY},Q_{X_0Y},\theta,R_{\mbox{\tiny w}})},
\end{equation}
where we have replaced $\hP_{\bx\by}$ by the notation $Q_{X_0Y}$ ($X_0$
being an auxiliary random variable that represents the underlying
source vector $\bx$), and where
\begin{equation}
E(R_{\mbox{\tiny w}},Q_{X_0Y},Q_{XY},\theta)=\left\{\begin{array}{ll}
[R_{\mbox{\tiny w}}-H_Q(X|Y)]_+ & \theta >
a(Q_{X_0Y})-a(Q_{XY})-[H_Q(X|Y)-R_{\mbox{\tiny w}}]_+\\
\infty & \theta \le a(Q_{X_0Y})-a(Q_{XY})-[H_Q(X|Y)-R_{\mbox{\tiny
w}}]_+\end{array}\right.
\end{equation}
Thus,
\begin{eqnarray}
\bar{P}_{\mbox{\tiny FR}}(\bx,\by,\bs,\bw)&\exe&
\max_{Q_{X|Y}}\int_{[a(Q_{X_0Y})-a(Q_{XY})-[H_Q(X|Y)-R_{\mbox{\tiny w}}]_+]_+}^\infty
\mbox{d}\theta e^{-n\theta}\cdot e^{-n[R_{\mbox{\tiny
w}}-H_Q(X|Y)]_+}.
\end{eqnarray}
whose exponential decay rate is
according to
\begin{eqnarray}
& &\min_{Q_{X|Y}}\{[a(Q_{X_0Y})-a(Q_{XY})-[H_Q(X|Y)-R_{\mbox{\tiny w}}]_+]_++
[R_{\mbox{\tiny w}}-H_Q(X|Y)]_+\}\nonumber\\
&=&\min_{Q_{X|Y}}\left\{\begin{array}{ll}
[R_{\mbox{\tiny w}}-H_Q(X|Y)+a(Q_{X_0Y})-a(Q_{XY})]_+ & H_Q(X|Y) >
R_{\mbox{\tiny w}}\\
R_{\mbox{\tiny w}}-H_Q(X|Y)+
[a(Q_{X_0 Y})-a(Q_{XY})]_+ & H_Q(X|Y)\le
R_{\mbox{\tiny w}}\end{array}\right.\nonumber\\
&=&\min_{Q_{X|Y}}[R_{\mbox{\tiny w}}-H_Q(X|Y)+[a(Q_{X_0
Y})-a(Q_{XY})]_+]_+\nonumber\\
&=&E(R_{\mbox{\tiny w}},Q_{X_0Y}).
\end{eqnarray}
The second to the last equality follows from the identity
$[u-v]_+=[[u]_+-v]_+$, holding whenever $v \ge 0$, which is applied to the
first line of the second expression with the assignments
$u=a(Q_{X_0Y})-a(Q_{XY})$ and $v=H_Q(X|Y)-R_{\mbox{\tiny w}}$
(see also \cite{SMF15}
as well as the text after eq.\
(11) of \cite{p187} for a very similar argument).
Since this exponential behavior, 
of $\bar{P}_{\mbox{\tiny FR}}(\bx,\by,\bs,\bw)$, is independent
of the particular realizations, $\bs$ and $\bw$, it holds also for the
expectation w.r.t.\ the randomness of $\bS$ and $\bW$, namely, it also
characterizes the exponential rate of
$\bar{P}_{\mbox{\tiny FR}}(\bx,\by)$. Finally, it readily follows from the
method of types \cite{CK11} that the expectation w.r.t.\
the randomness of $(\bX,\bY)$ decays according to the exponent
\begin{equation}
E_{\mbox{\tiny r}}^{\mbox{\tiny FR}}(R_{\mbox{\tiny
w}})=\min_{Q_{X_0Y}}\{D(Q_{X_0Y}\|P_{XY})+E(R_{\mbox{\tiny w}},Q_{X_0Y})\},
\end{equation}
which is as defined in (\ref{ErR}). This completes the proof of Theorem 1.
$\Box$

\subsection*{B. Expurgated Bound}

Our expurgated bound will be asserted for each type class, $\calT(Q_X)$, of
source vectors separately. As in channel coding, where expurgation is
associated with elimination of some `bad' codewords of 
a randomly generated code, here too, we might need to eliminate a small
fraction of bad source vectors from 
$\calT(Q_X)$, in order to guarantee a certain FR performance level for each one
of the remaining source vectors in $\calT(Q_X)$. One may wonder what would be
the justification for such an elimination of source vectors, as these are
generated by the source and given to us, 
and they are not under our control. Nonetheless, in the context
of biometric identification system described in Section III, 
where $\{\bx\}$ are the enrollment signals, 
there are at least two possible ways to justify
this elimination of a small fraction of the members of the type class.
\begin{enumerate}
\item In the enrollment stage, if the individual that subscribes to the system,
has generated a `forbidden' source vector $\bx$ (in the sense that 
has been eliminated in the
expurgation process),
s/he might be asked to kindly provide his/her biometric signal
once again, with the hope that this time a `legitimate' source vector will be
generated. The probability that this would happen is small in the first place,
provided that the fraction of
vectors eliminated from $\calT(Q_X)$ is small. The probability of bothering
the subscriber more than once with the request of a repeated measurement
is even much smaller.
\item Considering the fact that $\bx$ may be digitized with some precision 
(which is in line with the
finite alphabet assumption anyway), it is conceivable to think of the
enrollment data as having undergone a certain stage of vector quantization.
Once $\bx$ is thought of as an output of a vector quantizer, then not
necessarily every member of $\calT(Q_X)$ must be a legitimate codebook vector
in the first place. Among other things, one might rule out source vectors that
contribute a high FR probability.
\end{enumerate}

In order to present the expurgated exponent, a few additional definitions are needed.
For a given $Q_Y$, let us define
\begin{equation}
\label{alphadef}
\alpha(R_{\mbox{\tiny w}},Q_Y)\dfn\sup_{\{Q_{X|Y}:~H_Q(X|Y) > R_{\mbox{\tiny
w}}\}}[a(Q_{XY})+H_Q(X|Y)]-R_{\mbox{\tiny w}},
\end{equation}
\begin{equation}
\label{gammadef}
\gamma(Q_{XY})\dfn\max\{a(Q_{XY}),\alpha(R_{\mbox{\tiny w}},Q_Y)\},
\end{equation}
\begin{equation}
\label{Lambdadef}
\Lambda(Q_{XX'})\dfn\min_{Q_{Y|XX'}}\{\gamma(Q_{XY})-H_Q(Y|X,X')-\bE_Q\ln
P(Y|X)-a(Q_{X'Y})\},
\end{equation}
and for a given $Q_X$, define
\begin{equation}
E_{\mbox{\tiny ex}}^{\mbox{\tiny FR}}(R_{\mbox{\tiny w}},Q_X)=
\inf_{\{Q_{X'|X}:~H_Q(X'|X)\ge R_{\mbox{\tiny
w}}\}}\{\Lambda(Q_{XX'})-H_Q(X'|X)+R_{\mbox{\tiny w}}\}.
\end{equation}
Finally, let $P_{\mbox{\tiny FR}}(\calE|\bx)$ denote the FR probability
of a given enrollment encoder $\calE$, conditioned on the input source vector $\bX=\bx$.

\begin{theorem}
\label{expurg}
Consider the system configuration described in Section III and let
$\{\delta_n\}_{n\ge 1}$ be a positive sequence tending to zero 
such that $n\delta_n\to\infty$.
Then, there exists a code $\calE$ such that for every $Q_X$,
\begin{equation}
P_{\mbox{\tiny FR}}(\calE|\bx)\leq \exp\{-nE_{\mbox{\tiny ex}}^{\mbox{\tiny
FR}}(R_{\mbox{\tiny w}},Q_X)+o(n)\},
\end{equation}
for every $\bx\in\calT(Q_X)\setminus\calB(Q_X)$, where
$\calB(Q_X)$ is a certain subset of $\calT(Q_X)$, whose size does not exceed
$e^{-n\delta_n}|\calT(Q_X)|$.
\end{theorem}

A few points concerning Theorem \ref{expurg} should be discussed.

\noindent
1. It is interesting to note that the expression of 
$E_{\mbox{\tiny ex}}^{\mbox{\tiny FR}}(R_{\mbox{\tiny w}},Q_X)$
has some analogy to the Csisz\'ar--K\"orner--Marton (CKM) expurgated exponent
of channel coding \cite[p.\ 165, Problem 10.18]{CK11}. 
The term $\Lambda(Q_{XX'})$ plays the same role
as the expected Bhattacharyya distance in the CKM
expurgated exponent, whereas $H_Q(X'|X)$ is analogous to the coding rate $R$
in channel coding and $R_{\mbox{\tiny w}}$ is parallel to the empirical mutual
information between channel codewords.
Roughly speaking, the contribution of a single incorrect source vector $\bx'$
to the FR probability is about $\exp\{-n\Lambda(Q_{XX'})$ provided that
$(\bx,\bx')\in\calT(Q_{XX'})$ (the pairwise error event). This
probability should be multiplied by the typical number of such incorrect
source vectors within $\calT(Q_{X'|X}|\bx)$ that are encoded into the same given
helper message and hence may cause confusion. This number is of the exponential order
$\exp\{n[H_Q(X'|X)-R_{\mbox{\tiny w}}]\}$, provided that
$H_Q(X'|X)-R_{\mbox{\tiny w}} > 0$, and it vanishes otherwise.\\

\noindent
2. Note that in contrast to Theorem \ref{fr}, here we are no longer arguing
that the result is ensemble--tight. There is actually one step in the
derivation where exponential tightness might be compromised.
Specifically, in one of the steps of this analysis, the denominator of (\ref{gld}) is lower
bounded by a relatively simple single--letter bound that holds true for the vast
majority of encoders, $\{\calE\}$, in the ensemble. 
By doing this, possible gaps to these bounds may 
not be fully exploited, and we cannot rule out the possibility that 
this causes some loss of tightness. On the other hand, the derivation of the 
expurgated bound includes a certain
degree of freedom that does not exist in the random coding bound of Theorem
\ref{fr}, and upon
exploiting this degree of freedom, we obtain a result, which is at least as
strong as the random coding bound, and sometimes strictly so.\\

\noindent
3. The sequence $\delta_n$ tends to zero in order not to slow down the
exponential decay rate, but it is also required that $n\delta_n\to\infty$ in
order to guarantee that the set of `bad' source vectors, $\calB(Q_X)$, would be
merely a minority of $\calT(Q_X)$ for large $n$.\\

\noindent
4. We now show that for every $R_{\mbox{\tiny w}}$, 
the overall expurgated exponent (taking into account all
types, $\{Q_X\}$) 
cannot be worse than $E_{\mbox{\tiny
r}}^{\mbox{\tiny FR}}(R_{\mbox{\tiny w}})$, at least for the
metric $a(Q_{XY})=-\beta H_Q(X|Y)$, which was shown to be as good as the
optimal decoding metric in the ordinary random coding sense.
Note that is in contrast to the traditional expurgated exponent, which improves on
the random coding exponent only at a certain range of rates, but is inferior
to the random coding exponent elsewhere (see also \cite{p187}, where a similar
finding was observed for a particular numerical example).
For the above--mentioned choice of $a(Q_{XY})$,
one easily verifies that $\alpha(R_{\mbox{\tiny w}},Q_Y)=-\beta R_{\mbox{\tiny
w}}$ and $\gamma(Q_{XY})=-\beta\min\{H_Q(X|Y),R_{\mbox{\tiny w}}\}$,
and so,
\begin{eqnarray}
\Lambda(Q_{XX'})&=&\min_{Q_{Y|XX'}}\{\gamma(Q_{XY})-H_Q(Y|X,X')-\bE_Q\ln
P(Y|X)+\beta H_Q(X'|Y)\}\nonumber\\
&=&\min_{Q_{Y|XX'}}\{\beta[H_Q(X'|Y)-\min\{H_Q(X|Y),R_{\mbox{\tiny
w}}\}]+\nonumber\\
& &I_Q(X';Y|X)+D(Q_{Y|X}\|P_{Y|X}|Q_X)\}.
\end{eqnarray}
Upon optimizing $\beta$, we obtain
\begin{eqnarray}
\label{compare}
E_{\mbox{\tiny ex}}(R_{\mbox{\tiny w}},Q_X)&=&
\sup_{\beta\in\reals}\inf_{\{Q_{X'|X}:~H_Q(X'|X)\ge R_{\mbox{\tiny
w}}\}}\{\Lambda(Q_{XX'})-H_Q(X'|X)\}+R_{\mbox{\tiny
w}}\nonumber\\
&=&\sup_{\beta\in\reals }\inf_{\{Q_{X'Y|X}:~H_Q(X'|X)\ge 
R_{\mbox{\tiny w}}\}}\{D(Q_{Y|X}\|P_{Y|X}|Q_X)+I_Q(X';Y|X)+\nonumber\\
& &\beta[H_Q(X'|Y)-\min\{H_Q(X|Y),R_{\mbox{\tiny
w}}\}]-
H_Q(X'|X)+R_{\mbox{\tiny
w}}\}\nonumber\\
&\ge&\inf_{\{Q_{X'Y|X}:~H_Q(X'|X)\ge 
R_{\mbox{\tiny w}}\}}\{D(Q_{Y|X}\|P_{Y|X}|Q_X)+I_Q(X';Y|X)+\nonumber\\
& &H_Q(X'|Y)-\min\{H_Q(X|Y),R_{\mbox{\tiny w}}\}-
H_Q(X'|X)+R_{\mbox{\tiny w}}\}\nonumber\\
&=&\inf_{\{Q_{X'Y|X}:~H_Q(X'|X)\ge 
R_{\mbox{\tiny w}}\}}\{D(Q_{Y|X}\|P_{Y|X}|Q_X)+
I_Q(X';Y|X)+H_Q(X'|Y)+\nonumber\\
& &[R_{\mbox{\tiny
w}}-H_Q(X|Y)]_+-
H_Q(X'|X)\nonumber\\
&=&\inf_{\{Q_{X'Y|X}:~H_Q(X'|X)\ge 
R_{\mbox{\tiny w}}\}}\{D(Q_{Y|X}\|P_{Y|X}|Q_X)+
H_Q(X'|Y)-H_Q(X'|X,Y)+\nonumber\\
& &[R_{\mbox{\tiny
w}}-H_Q(X|Y)]_+\nonumber\\
&=&\inf_{\{Q_{X'Y|X}:~H_Q(X'|X)\ge R_{\mbox{\tiny
w}}\}}\{D(Q_{Y|X}\|P_{Y|X}|Q_X)+I_Q(X';X|Y)+[R_{\mbox{\tiny
w}}-H_Q(X|Y)]_+\}\nonumber\\
&\ge&\inf_{\{Q_{X'Y|X}:~H_Q(X'|X)\ge R_{\mbox{\tiny
w}}\}}\{D(Q_{Y|X}\|P_{Y|X}|Q_X)+[R_{\mbox{\tiny w}}-H_Q(X|Y)]_+\}.
\end{eqnarray}
Without the constraint, $H_Q(X'|X)\ge R_{\mbox{\tiny w}}$, 
the last expression is exactly the random coding FR exponent for a given type
$Q_X$, and upon taking into account the probabilistic weight of each type, the
overall exponent associated with the last line (again, without the constraint) 
is exactly $E_{\mbox{\tiny
r}}(R_{\mbox{\tiny m}})$ of Theorem \ref{fr} for the optimal, MAP decoder.
By inspection of eq.\ (\ref{compare}), 
we therefore observe that there are four origins of the gap between the expurgated
exponent and the random coding exponent:
(i) the decoder actually being analyzed might be
suboptimal for the expurgated ensemble, 
(ii) the optimal $\beta$ (for the 
given family of decoders) might not necessarily be
$\beta^*=1$ (the first inequality in the above chain).
In fact, the optimal $\beta^*$ is expected to depend on
$R_{\mbox{\tiny w}}$.\footnote{The fact that optimal $\beta$ may not necessarily
be infinite (except 
the case (\ref{ordinary2})), is interesting on its own right, as it means that
the the stochastic decoder may outperform the deterministic one for a given
(suboptimal) decoding metric.}
(iii) the term $I_Q(X';X|Y)$ which may not necessarily
vanish for the optimal $Q_{X'Y|X}$ (the second inequality),
and (iv) the constraint $H_Q(X'|X)\ge
R_{\mbox{\tiny w}}$.
For example, if $R_{\mbox{\tiny w}}> \ln|\calX|$, the
expurgated exponent is infinite while the random coding exponent is finite.\\

\noindent
5. As can be seen in the proof of Theorem \ref{expurg}, the asserted
expurgated exponent is obtained from an intermediate expression that depends
on a free parameter $\rho$ that undergoes optimization. 
It is interesting to observe what happens when we set $\rho=1$
instead of optimizing over $\rho$. This would correspond to the ordinary
ensemble average, which needs no expurgation. In this case, $E_{\mbox{\tiny
ex}}^{\mbox{\tiny
FR}}(R_{\mbox{\tiny w}},Q_X)$ would
be replaced by
\begin{eqnarray}
E_1(R_{\mbox{\tiny w}},Q_X)&=&\sup_{\beta\in\reals}
\inf_{Q_{X'|X}}\left\{\Lambda(Q_{XX'})-[H_Q(X'|X)-R_{\mbox{\tiny\mbox
w}}]_++[R_{\mbox{\tiny w}}-H_Q(X'|X)]_+\right\}\nonumber\\
&=&\sup_{\beta\in\reals}\inf_{Q_{X'|X}}\left\{\Lambda(Q_{XX'})+
R_{\mbox{\tiny w}}-H_Q(X'|X)\right\},
\end{eqnarray}
where we have used the trivial identity $[u]_+-[-u]_+\equiv u$.
Therefore, the expression of $E_1(R_{\mbox{\tiny w}},Q_X)$ is exactly like
that of $E_{\mbox{\tiny ex}}^{\mbox{\tiny FR}}(R_{\mbox{\tiny m}},Q_X)$,
except that the constraint, $H_Q(X'|X)\ge
R_{\mbox{\tiny w}}$, is removed. It follows that $E_{\mbox{\tiny
ex}}^{\mbox{\tiny FR}}(R_{\mbox{\tiny
w}},Q_X)$ is expected to improve on $E_1(R_{\mbox{\tiny w}},Q_X)$ at
high rates, where the constraint may be active.
It also follows (similarly as in (\ref{compare}))
that $E_1(R_{\mbox{\tiny w}},Q_X)$ is never smaller than the random coding
FR exponent given the type $Q_X$,
since the latter lacks this constraint as well.
The reason that this expurgated exponent is nowhere worse than the random
coding exponent is that we do not use the inequality $[\sum_{\bx'\ne\bx}
u(\bx')]^{1/\rho}\leq \sum_{\bx'\ne\bx}[u(\bx')]^{1/\rho}$ (holding for
$\rho\ge 1$), like in the
traditional expurgated bound. This inequality causes a loss of tightness.
Without it, the supremum over $\rho$ 
is always achieved at $\rho\to\infty$.\\

\noindent
6. The case of ordinary, deterministic MAP decoding 
is obtained again as of special case
of (\ref{ordinary2}) in
the limit $\beta\to\infty$. As in (\ref{maplimit}), when the objective function to be minimized
over $\{Q_{XX'Y})\}$, contains a term like $\beta\cdot
G(Q_{XX'Y})$ (for some functional $G(\cdot)$), then in the limit of
$\beta\to\infty$, it is replaced by
a constraint of the form
$G(Q_{XX'Y})\le 0$.\\

The remaining part of this section is devoted to the proof of Theorem
\ref{expurg}.

\noindent
{\it Proof of Theorem \ref{expurg}.}
For a given code, $\calE$, and a given the underlying source
vector $\bx$, we have
\begin{eqnarray}
P_{\mbox{\tiny FR}}(\calE|\bx)
&=&\sum_{\by}P(\by|\bx)\sum_{\bs\ne
g(\bx)}\tilde{P}(\bs|f(\bx),\by)\\
&=&\sum_{\bs\ne g(\bx)}
\sum_{\by}P(\by|\bx)\cdot\frac{\tilde{P}(\bs,f(\bx)|\by)}
{\exp\{na(\hat{P}_{\bx\by})\}+Z_{\bx}(\by)},
\end{eqnarray}
where 
\begin{equation}
Z_{\bx}(\by)=\sum_{\bx^\prime\ne\bx}\exp\{na(\hP_{\bx^\prime\by})\}
\cdot\calI\{f(\bx^\prime)=
f(\bx)\}.
\end{equation}
Let $\epsilon > 0$ be arbitrarily small.
It is shown in the Appendix\footnote{See also
\cite[Appendix B]{p187} for a similar argument related to channel coding.}
that 
\begin{equation}
\label{appendixb}
\mbox{Pr}\left\{Z_{\bx}(\by) <
\exp\{n\alpha(R_{\mbox{\tiny w}}+\epsilon,\hP_{\by})\}~\mbox{for some}~(\bx,\by)\right\}\le
|\calX\times\calY|^n\cdot\exp\{-e^{n\epsilon}+n\epsilon+1\}.
\end{equation}
Now, denoting
\begin{equation}
\calG_\epsilon=\left\{\calE:~Z_{\bx}(\by) \ge
\exp\{n\alpha(R_{\mbox{\tiny w}}+\epsilon,\hP_{\by})\}~\mbox{for
all}~(\bx,\by)\right\},
\end{equation}
we have:
\begin{eqnarray}
& &\bE\left\{[P_{\mbox{\tiny FR}}(\calE|\bx)]^{1/\rho}\right\}\nonumber\\
&=&\bE\left[\sum_{\bs\ne g(\bx)}
\sum_{\by}P(\by|\bx)\cdot\frac{\tilde{P}(\bs,f(\bx)|\by)}
{\exp\{na(\hat{P}_{\bx\by})\}+Z_{\bx}(\by)}\right]^{1/\rho}\nonumber\\
&=&\sum_{\calE}P(\calE)\left[\sum_{\bs\ne g(\bx)}
\sum_{\by}P(\by|\bx)\cdot\frac{\tilde{P}(\bs,f(\bx)|\by)}
{\exp\{na(\hat{P}_{\bx\by})\}+Z_{\bx}(\by)}\right]^{1/\rho}\nonumber\\
&=&\sum_{\calE\in\calG_\epsilon}P(\calE)\left[\sum_{\bs\ne g(\bx)}
\sum_{\by}P(\by|\bx)\cdot\frac{\tilde{P}(\bs,f(\bx)|\by)}
{\exp\{na(\hat{P}_{\bx\by})\}+Z_{\bx}(\by)}\right]^{1/\rho}+\nonumber\\
& &\sum_{\calE\in\calG_\epsilon^c}P(\calE)\left[\sum_{\bs\ne g(\bx)}
\sum_{\by}P(\by|\bx)\cdot\frac{\tilde{P}(\bs,f(\bx)|\by)}
{\exp\{na(\hat{P}_{\bx\by})\}+Z_{\bx}(\by)}\right]^{1/\rho}\nonumber\\
&\le&\sum_{\calE\in\calG_\epsilon}P(\calE)\left[\sum_{\bs\ne g(\bx)}
\sum_{\by}P(\by|\bx)\cdot
\frac{\tilde{P}(\bs,f(\bx)|\by)}
{\exp\{na(\hat{P}_{\bx\by})\}+
\exp\{n\alpha(R_{\mbox{\tiny w}}+\epsilon,\hat{P}_{\by})\}}\right]^{1/\rho}+\nonumber\\
& &\sum_{\calE\in\calG_\epsilon^c}P(\calE)\cdot
1^{1/\rho}\nonumber\\
&\le&\sum_{\calE}P(\calE)\left[\sum_{\bs\ne g(\bx)}
\sum_{\by}P(\by|\bx)\cdot\frac{\tilde{P}(\bs,f(\bx)|\by)}
{\exp\{na(\hat{P}_{\bx\by})\}+
\exp\{n\alpha(R_{\mbox{\tiny w}}+\epsilon,\hat{P}_{\by})\}}\right]^{1/\rho}+\nonumber\\
& & e^{nR_{\mbox{\tiny s}}}\cdot|\calX\times\calY|^n\cdot\exp\{-e^{n\epsilon}+n\epsilon+1\}.
\end{eqnarray}
Considering the arbitrariness of $\epsilon$,
the expression in the square brackets is exponentially equivalent to
\begin{eqnarray}
& &\sum_{\bs\ne g(\bx)}
\sum_{\by}P(\by|\bx)e^{-n\gamma(\hat{P}_{\bx\by})} \tilde{P}(\bs,f(\bx)|\by)\nonumber\\
&=&\sum_{\bs\ne g(\bx)}
\sum_{\by}P(\by|\bx)e^{-n\gamma(\hat{P}_{\bx\by})}
\sum_{\bx'}\exp\{na(\hP_{\bx'\by})\}\calI\{f(\bx')=f(\bx),g(\bx')=\bs\}\nonumber\\
&=&\sum_{\bs\ne g(\bx)}
\sum_{\bx'}\calI\{f(\bx')=f(\bx),g(\bx')=\bs\}
\sum_{\by}P(\by|\bx)\exp\{n[a(\hP_{\bx'\by})-\gamma(\hat{P}_{\bx\by})]\}.
\end{eqnarray}
Now, the inner most summation (over $\by$) can be assessed using the method of
types \cite{CK11}. Accordingly, referring to (\ref{Lambdadef}), we have
\begin{equation}
e^{-n\Lambda(\hat{P}_{\bx\bx'})}\exe\sum_{\by}P(\by|\bx)\exp\{n[a(\hP_{\bx'\by})-
\gamma(\hat{P}_{\bx\by})]\},
\end{equation}
which is the contribution of a single incorrect source vector
$\bx'$ to the FR probability.
This yields
\begin{eqnarray}
& &\sum_{\bs\ne g(\bx)}
\sum_{\bx'}\calI\{f(\bx')=f(\bx),g(\bx')=\bs\}\cdot
e^{-n\Lambda(\hat{P}_{\bx\bx'})}\nonumber\\
&\le&\sum_{\bx'}
e^{-n\Lambda(\hat{P}_{\bx\bx'})}\calI\{f(\bx')=f(\bx)\}\nonumber\\
&=&\sum_{Q_{X'|X}}
e^{-n\Lambda(Q_{XX'})}N(\calT(Q_{X'|X}|\bx),f(\bx)),
\end{eqnarray}
where we have defined
\begin{equation}
N(\calT(Q_{X'|X}|\bx),f(\bx))\dfn
\bigg|\calT(Q_{X'|X}|\bx)\cap\{\bx':~f(\bx')=f(\bx)\}\bigg|.
\end{equation}
On substituting this back into the bound on $\bE\left\{[P_{\mbox{\tiny
FR}}(\calE|\bx)]^{1/\rho}\right\}$, we get
\begin{eqnarray}
& &\bE\left\{[P_{\mbox{\tiny e}}(\calE|\bx)]^{1/\rho}\right\}\nonumber\\
&\le&\bE\left\{\left[\sum_{Q_{X'|X}}e^{-n\Lambda(Q_{XX'})}
N(\calT(Q_{X'|X}|\bx),f(\bx))
\right]^{1/\rho}\right\}\nonumber\\
&\exe&\sum_{Q_{X'|X}}e^{-n\Lambda(Q_{XX'})/\rho}\bE\left\{
[N(\calT(Q_{X'|X}|\bx),f(\bx))]^{1/\rho}\right\}\nonumber\\
&=&\sum_{Q_{X'|X}}e^{-n\Lambda(Q_{XX'})/\rho}\int_0^\infty\mbox{d}t\cdot
\mbox{Pr}\left\{\left[
N(\calT(Q_{X'|X}|\bx),f(\bx))\right]^{1/\rho}\ge t\right\}\nonumber\\
&=&\sum_{Q_{X'|X}}e^{-n\Lambda(Q_{XX'})/\rho}\int_0^\infty\mbox{d}t\cdot
\mbox{Pr}\left\{
N(\calT(Q_{X'|X}|\bx),f(\bx))
\ge t^\rho\right\}\nonumber\\
&\exe&\sum_{Q_{X'|X}}e^{-n\Lambda(Q_{XX'})/\rho}\int_{-\infty}^\infty\mbox{d}\theta\cdot
e^{n\theta}\cdot
\mbox{Pr}\left\{N(\calT(Q_{X'|X}|\bx),f(\bx))
\ge e^{n\theta\rho}\right\}.
\end{eqnarray}
Let us focus on the term $\mbox{Pr}[N(\calT(Q_{X'|X}|\bx),f(\bx))\ge
e^{n\theta\rho}]$. Since $N(\calT(Q_{X'|X}|\bx),f(\bx))$ is a binomial random
variable with $|\calT(Q_{X'|X}|\bx)|\exe e^{nH_Q(X'|X)}$ trials and probability of success
$e^{-nR_{\mbox{\tiny w}}}$, we have
\begin{equation}
\mbox{Pr}\left[N(\calT(Q_{X'|X}|\bx),f(\bx))\ge
e^{n\theta\rho}\right]
\exe e^{-nE(R_{\mbox{\tiny w}},Q_{XX'},\rho\theta)}
\end{equation}
where
\begin{eqnarray}
E(R_{\mbox{\tiny w}},Q_{XX'},\rho\theta)&=&\left\{\begin{array}{ll}
[R_{\mbox{\tiny w}}-H_Q(X'|X)]_+ & [H_Q(X'|X)-R_{\mbox{\tiny w}}]_+ \ge
\rho\theta\\
\infty & [H_Q(X'|X)-R_{\mbox{\tiny w}}]_+<\rho\theta
\end{array}\right.\nonumber\\
&=&\left\{\begin{array}{ll}
[R_{\mbox{\tiny w}}-H_Q(X'|X)]_+ & \theta \le [H_Q(X'|X)-R_{\mbox{\tiny
w}}]_+/\rho\\
\infty & \theta > [H_Q(X'|X)-R_{\mbox{\tiny
w}}]_+/\rho\end{array}\right.
\end{eqnarray}
On substituting this back into the expression of 
$\bE\left\{[P_{\mbox{\tiny
FR}}(\calE|\bx)]^{1/\rho}\right\}$, we get
\begin{eqnarray}
& &\bE\left\{[P_{\mbox{\tiny FR}}(\calE|\bx)]^{1/\rho}\right\}\nonumber\\
&\lexe&\sum_{Q_{X'|X}}e^{-n\Lambda(Q_{XX'})/\rho}\cdot
\int_{-\infty}^{[H_Q(X'|X)-R_{\mbox{\tiny\mbox w}}]_+/\rho}{d}\theta\cdot
e^{n\theta}e^{-n[R_{\mbox{\tiny w}}-H_Q(X'|X)]_+}\nonumber\\
&\exe&\exp\left\{-n\min_{Q_{X'|X}}\left[\Lambda(Q_{XX'})
+\rho[R_{\mbox{\tiny w}}-H_Q(X'|X)]_+-[H_Q(X'|X)-R_{\mbox{\tiny\mbox
w}}]_+\right]/\rho\right\}\nonumber\\
&\dfn&e^{-nE_{\mbox{\tiny x}}(R_{\mbox{\tiny w}},Q_X,\rho)/\rho}.
\end{eqnarray}
It follows then that
\begin{equation}
\bE\left\{\frac{1}{|\calT(Q_X)|}\sum_{\bx\in\calT(Q_X)}[P_{\mbox{\tiny
FR}}(\calE|\bx)]^{1/\rho}\right\}\lexe
e^{-nE_{\mbox{\tiny x}}(R_{\mbox{\tiny w}},Q_X,\rho)/\rho},
\end{equation}
and so, there exists a code $\calE$ with
\begin{equation}
\frac{1}{|\calT(Q_X)|}\sum_{\bx\in\calT(Q_X)}[P_{\mbox{\tiny
FR}}(\calE|\bx)]^{1/\rho}\lexe
e^{-nE_{\mbox{\tiny x}}(R_{\mbox{\tiny w}},Q_X,\rho)/\rho}.
\end{equation}
For a given such $\calE$ and $Q_X$, let us order the members of $\calT(Q_X)$,
as $\bx_1,\bx_2,\bx_3,\ldots$,
according to $P_{\mbox{\tiny FR}}(\calE|\bx_1)\ge P_{\mbox{\tiny
FR}}(\calE|\bx_2)\ge P_{\mbox{\tiny FR}}(\calE|\bx_3)\ge\ldots$ and let
$M$ be a temporary short-hand notation for $|\calT(Q_X)|$. Let
$\calB(Q_X)$ be the subset of $\calT(Q_X)$ formed by the first
$M^\prime=e^{-\delta n}M$ members of $\calT(Q_X)$ according to this order,
i.e., $B(Q_X)=
\{\bx_1,\bx_2,\ldots,\bx_{M^\prime}\}$. We then have
\begin{eqnarray}
e^{-nE_{\mbox{\tiny x}}(R_{\mbox{\tiny w}},Q_X,\rho)/\rho}&\gexe&
\frac{1}{M}\sum_{m=1}^M[P_{\mbox{\tiny FR}}(\calE|\bx_m)]^{1/\rho}\nonumber\\
&\ge&\frac{1}{M}\sum_{m=1}^{M^\prime}[P_{\mbox{\tiny FR}}(\calE|\bx_m)]^{1/\rho}\nonumber\\
&\ge&\frac{1}{M}\sum_{m=1}^{M^\prime}[P_{\mbox{\tiny
FR}}(\calE|\bx_{M^\prime+1})]^{1/\rho}\nonumber\\
&=&\frac{1}{M}\cdot M^\prime\cdot [P_{\mbox{\tiny
FR}}(\calE|\bx_{M^\prime+1})]^{1/\rho}\nonumber\\
&=&e^{-n\delta_n}\left[\max_{\bx\in\calT(Q_X)\setminus
B(Q_X)}P_{\mbox{\tiny FR}}(\calE|\bx)\right]^{1/\rho},
\end{eqnarray}
and so,
$\max_{\bx\in\calT(Q_X)\setminus B(Q_X)}P_{\mbox{\tiny FR}}(\calE|\bx)$
decays at an exponential rate which is at least as large as
\begin{eqnarray}
\label{Eex}
& &\sup_{\rho\ge 0}E_{\mbox{\tiny x}}(R_{\mbox{\tiny w}},Q_X,\rho)\nonumber\\
&=&\sup_{\rho\ge
0}\inf_{Q_{X'|X}}\left\{\Lambda(Q_{XX'})-[H_Q(X'|X)-R_{\mbox{\tiny\mbox
w}}]_++\rho[R_{\mbox{\tiny w}}-H_Q(X'|X)]_+\right\}\\
&=&\inf_{\{Q_{X'|X}:~H_Q(X'|X)\ge R_{\mbox{\tiny
w}}\}}\left\{\Lambda(Q_{XX'})-[H_Q(X'|X)-R_{\mbox{\tiny
w}}]_+\right\}\nonumber\\
&=&\inf_{\{Q_{X'|X}:~H_Q(X'|X)\ge R_{\mbox{\tiny
w}}\}}\{\Lambda(Q_{XX'})-H_Q(X'|X)+R_{\mbox{\tiny w}}\}\nonumber\\
&=&E_{\mbox{\tiny x}}(R_{\mbox{\tiny w}},Q_X),
\end{eqnarray}
completing the proof of Theorem \ref{expurg}.

\section*{V. False--Accept Error Analysis}

In this section, we analyze the ensemble performance of the system from the
viewpoint of an imposter who makes an attempt to estimate the secret key without access
to the side information $\bY$, and we are interested in the exponential decay
rate of the FA probability for the average code. As described in Section III,
here we assume that the imposter estimates $\bS$ using the MAP estimator,
$\tilde{\bS}$ (see (\ref{imposter})), based
on the helper message only. Accordingly, as defined in Section III, we denote
$\bar{P}_{\mbox{\tiny FA}}=\mbox{Pr}\{\tilde{\bS}=\bS\}$, i.e., the
probability of correct decoding (FA), averaged over the ensemble of codes
$\{\calE\}$. Let us define
\begin{equation}
E_{\mbox{\tiny FA}}(R_{\mbox{\tiny w}},R_{\mbox{\tiny s}})=\min_{Q_X}\left[D(Q_X\|P_X)+\min\{
R_{\mbox{\tiny s}},[H_Q(X)-R_{\mbox{\tiny w}}]_+\}\right].
\end{equation}

Our main result, in this section, is the following.

\begin{theorem}
\label{fa}
Consider the system configuration described in Section III. Then,
\begin{equation}
\bar{P}_{\mbox{\tiny FA}}\le \exp\{-nE_{\mbox{\tiny FA}}(R_{\mbox{\tiny
w}},R_{\mbox{\tiny s}})+o(n)\}.
\end{equation}
\end{theorem}

The expression of this exponential error bound is quite intuitive and
it can easily be understood to hold even if the
imposter is informed about the type\footnote{Here a genie--aided
decoding argument does not harm the tightness of the FA exponent, 
because one can guess the type correctly
with probability of success that decays only polynomially.}
$Q_X$ of $\bX$.
There are about $e^{n[H_Q(X)-R_{\mbox{\tiny w}}]_+}$
source sequences
of type $Q_X$ (including the correct one), whose helper message is the given $\bW$. If
$[H_Q(X)-R_{\mbox{\tiny w}}]_+ > R_{\mbox{\tiny s}}$,
then all possible $e^{nR_{\mbox{\tiny s}}}$ members of the secret--message set 
would be likely to appear 
as encoded secret messages among those
sequences, approximately evenly, so the
probability of guessing the correct one is about $e^{-nR_{\mbox{\tiny s}}}$.
If, on the other hand, $[H_Q(X)-R_{\mbox{\tiny w}}]_+ < R_{\mbox{\tiny s}}$, then
it is very likely that there would be only about 
$e^{n[H_Q(X)-R_{\mbox{\tiny w}}]_+}$
different $\bs$--messages, so the probability of guessing
the correct one is the reciprocal, $e^{-n[H_Q(X)-R_{\mbox{\tiny w}}]_+}$.
It is easy to see that $E_{\mbox{\tiny FA}}(R_{\mbox{\tiny w}},R_{\mbox{\tiny s}})$
vanishes for $R_{\mbox{\tiny w}} > H(X)$, as expected.

It is also interesting to observe that here, in contrast to the exponential FR bounds of
Section IV, the exponent depends on both $R_{\mbox{\tiny w}}$ and
$R_{\mbox{\tiny s}}$, and not only on $R_{\mbox{\tiny w}}$.
As expected, it is increasing in $R_{\mbox{\tiny s}}$ and decreasing in
$R_{\mbox{\tiny w}}$.

The FA error exponent of Theorem \ref{fa} can also be presented in a
Gallager--style form:
\begin{eqnarray}
E_{\mbox{\tiny FA}}(R_{\mbox{\tiny
w}},R_{\mbox{\tiny s}})
&=&\min_Q[D(Q_X\|P_X)+\min\{R_{\mbox{\tiny s}},[H_Q(X)-R_{\mbox{\tiny
w}}]_+\}]\nonumber\\
&=&\min_{Q_X}\min_{0\le s\le 1}\max_{0\le \rho\le 1}\{D(Q_X\|P_X)+sR_{\mbox{\tiny s}}+
(1-s)\rho[H_Q(X)-R_{\mbox{\tiny w}}]\}\nonumber\\
&=&\min_{0\le s\le 1}\max_{0\le \rho\le 1}\min_{Q_X}\{D(Q_X\|P_X)+sR_{\mbox{\tiny s}}+
(1-s)\rho[H_Q(X)-R_{\mbox{\tiny w}}]\}\nonumber\\
&=&\min_{0\le s\le 1}\max_{0\le \rho\le 1}\left\{-[1-\rho(1-s)]\ln\left[\sum_x
P_X(x)^{1/[1-\rho(1-s)]}\right]+sR_{\mbox{\tiny s}}-\rho(1-s)R_{\mbox{\tiny
w}}\right\}\nonumber\\
&=&\min_{0\le s\le 1}\max_{s\le \rho\le 1}\left\{-\rho\ln\left[\sum_x
P_X(x)^{1/\rho}\right]+sR_{\mbox{\tiny s}}-(1-\rho)R_{\mbox{\tiny
w}}\right\}.
\end{eqnarray}

\noindent
{\it Proof of Theorem \ref{fa}.}
In the derivation below, we let $\bx_Q$ denote an arbitrary representative source vector
$\bx$ of type $Q_X$. The choice of this representative within $\calT(Q_X)$ is
completely immaterial since all members of $\calT(Q_X)$ are equiprobable.
Similarly as before, we also denote by $N(Q_X,\bw,\bs)$
the number of members of $\calT(Q_X)$ that are encoded into $(\bw,\bs)$.
\begin{eqnarray}
\bar{P}_{\mbox{\tiny FA}}&=&\bE\left\{\sum_{\bw}\max_{\bs}
P(\bw,\bs)\right\}\nonumber\\
&=&\bE\left\{\sum_{\bw}\max_{\bs}
\sum_{Q_X} P_X(\bx_Q)\cdot N(Q_X,\bw,\bs)\right\}\nonumber\\
&=&\lim_{\beta\to\infty}\bE\left\{\sum_{\bw}\left[\sum_{\bs}
\left(\sum_{Q_X} P_X(\bx_Q)\cdot N(Q_X,\bw,\bs)\right)^\beta\right]^{1/\beta}\right\}\nonumber\\
&\exe&\lim_{\beta\to\infty}\bE\left\{\sum_{\bw}\left[\sum_{\bs}
\sum_{Q_X} P_X^\beta(\bx_Q)\cdot N^\beta(Q_X,\bw,\bs)\right]^{1/\beta}\right\}\nonumber\\
&=&\lim_{\beta\to\infty}\bE\left\{\sum_{\bw}\left[\sum_{Q_X}
\sum_{\bs} P_X^\beta(\bx_Q)\cdot N^\beta(Q_X,\bw,\bs)\right]^{1/\beta}\right\}\nonumber\\
&\exe&\lim_{\beta\to\infty}\bE\left\{\sum_{\bw}\sum_{Q_X}
\left[\sum_{\bs} P_X^\beta(\bx_Q)\cdot
N^\beta(Q_X,\bw,\bs)\right]^{1/\beta}\right\}\nonumber\\
&=&\lim_{\beta\to\infty}\bE\left\{\sum_{\bw}\sum_{Q_X}P_X(\bx_Q)
\left[\sum_{\bs}N^\beta(Q_X,\bw,\bs)\right]^{1/\beta}\right\}\nonumber\\
&=&\lim_{\beta\to\infty}\sum_{\bw}\sum_{Q_X}P_X(\bx_Q)\cdot\bE\left\{
\left[\sum_{\bs}N^\beta(Q_X,\bw,\bs)\right]^{1/\beta}\right\}\nonumber\\
&=&\sum_{\bw}\sum_{Q_X}P_X(\bx_Q)\cdot\bE\left\{\max_{\bs}N(Q_X,\bw,\bs)\right\}\nonumber\\
&=&\sum_{\bw}\sum_{Q_X}P_X(\bx_Q)\cdot\sum_{n=1}^{|\calT(Q_X)|}
\mbox{Pr}\left\{\max_{\bs}N(Q_X,\bw,\bs)\ge n\right\}\nonumber\\
&=&\sum_{\bw}\sum_{Q_X}P_X(\bx_Q)\cdot\sum_{n=1}^{|\calT(Q_X)|}
\mbox{Pr}\bigcup_{\bs}\left\{N(Q_X,\bw,\bs)\ge n\right\}\nonumber\\
&\le&\sum_{\bw}\sum_{Q_X}P_X(\bx_Q)\cdot\sum_{n=1}^{|\calT(Q_X)|}
\min\left\{1, e^{nR_{\mbox{\tiny s}}}\mbox{Pr}\left[N(Q_X,\bw,\bs)\ge
n\right]\right\}.
\end{eqnarray}
Now, for $Q_X\in\calG\dfn\{Q_X:~H_Q(X) > R_{\mbox{\tiny s}}+R_{\mbox{\tiny w}}\}$,
clearly, $\mbox{Pr}[N(Q_X,\bw,\bs)\ge n]$ is large for every
$n\le e^{n[H_Q(X)-R_{\mbox{\tiny w}}-R_{\mbox{\tiny s}}-\epsilon]}$
(for an arbitrarily small $\epsilon > 0$ and large $n$),
and so, the
minimum between $1$ and $e^{nR_{\mbox{\tiny s}}}\mbox{Pr}\left[N(Q_X,\bw,\bs)\ge
n\right]$ is certainly 1. Hence, these terms, of the summation over $n$, 
contribute altogether a quantity of the exponential order of
$e^{n[H_Q(X)-R_{\mbox{\tiny w}}-R_{\mbox{\tiny s}}]}$. For larger $n$,
$\mbox{Pr}[N(Q_X,\bw,\bs)\ge n]$ decays super--exponentially, and so, these
terms contribute a negligible amount. Consequently, considering the factor of
$e^{nR_{\mbox{\tiny w}}}$ that stems from the summation over $\bw$, one term that
contributes to the expression of the last line above is
$\sum_{Q_X\in\calG}P_X(\bx_Q)e^{n[H_Q(X)-R_{\mbox{\tiny s}}]}$,
which is of the exponential order of
$\exp\{-n\min_{Q_X\in\calG}[D(Q_X\|P_X)+R_{\mbox{\tiny s}}]\}$. The other term
comes from the types that belong to $\calG^c$.
For $Q_X\in\calG^c$, there are sub--exponentially few terms that contribute
$\min\{1,e^{nR_{\mbox{\tiny s}}}\cdot e^{n[H_Q(X)-R_{\mbox{\tiny
s}}-R_{\mbox{\tiny w}}]}\}=e^{-n[R_{\mbox{\tiny w}}-H_Q(X)]_+}$, and so, the
overall contribution is $\max_{Q_X\in\calG^c}e^{nR_{\mbox{\tiny
w}}}e^{-n[H_Q(X)+D(Q_X\|P_X)]}e^{-n[R_{\mbox{\tiny w}}-H_Q(X)]_+}$, which is
$\exp\{-n\min_{Q_X\in\calG^c}[D(Q_X\|P_X)+[H_Q(X)-R_{\mbox{\tiny w}}]_+]\}$.
Thus, the overall performance is
\begin{equation}
\bar{P}_{\mbox{\tiny FA}}\lexe\exp\left(-n\min_{Q_X}[D(Q_X\|P_X)+\min\{R_{\mbox{\tiny
s}},[H_Q(X)-R_{\mbox{\tiny w}}]_+\}]\right), 
\end{equation}
completing the proof of Theorem \ref{fa}.

\section*{VI. Information Leakage for the Typical Code}

In this last section, which is very brief, we provide an outline for the evaluation of 
the third figure of merit of our model of an
authentication system, namely, the secrecy, or the information leakage,
$I(\bW;\bS)$, associated with the typical code, $\calE$, in the ensemble.

We envision the typical code as a code with the following properties: 
\begin{enumerate}
\item For any given type class
$\calT(Q_X)$ whose size is larger than $e^{n(R_{\mbox{\tiny s}}+R_{\mbox{\tiny
w}})}$, the number of members of $\calT(Q_X)$ mapped each one of the $e^{n(R_{\mbox{\tiny
s}}+R_{\mbox{\tiny w}})}$ pairs $(\bs,\bw)$ is exactly the same (uniform
distribution of $(\bS,\bW)$ within the type), so that
$H(\bS,\bW|\bX\in\calT(Q_X))=n(R_{\mbox{\tiny s}}+R_{\mbox{\tiny w}})$.
\item For any given type class
$\calT(Q_X)$ whose size is smaller than $e^{n(R_{\mbox{\tiny s}}+R_{\mbox{\tiny
w}})}$, each member of $\calT(Q_X)$ is mapped to a different pair $(\bs,\bw)$,
so that
$H(\bS,\bW|\bX\in\calT(Q_X))=\log|\calT(Q_X)|$.
\end{enumerate}

The leakage will then be upper bounded as follows:
\begin{eqnarray}
I(\bS;\bW)&=&H(\bS)+H(\bW)-H(\bS,\bW)\nonumber\\
&\le&nR_{\mbox{\tiny s}}+nR_{\mbox{\tiny w}}-H(\bS,\bW|\hP_{\bX})\nonumber\\
&=&n(R_{\mbox{\tiny s}}+R_{\mbox{\tiny w}})-\bE\min\left\{n(R_{\mbox{\tiny
s}}+R_{\mbox{\tiny w}}),\log|\calT(\hP_{\bX})|\right\}\nonumber\\
&=&\bE\left\{\left[n(R_{\mbox{\tiny s}}+R_{\mbox{\tiny
w}})-\log|\calT(\hP_{\bX})|\right]_+\right\}\nonumber\\
&\approx&n\bE\left\{[R_{\mbox{\tiny s}}+R_{\mbox{\tiny w}}-\hH_{\bX}(X)]_+\right\}.
\end{eqnarray}
Now, assuming that $H(X) > R_{\mbox{\tiny s}}+R_{\mbox{\tiny w}}$, the
probability of falling in a type class $\calT(\hP_{\bx})$ 
with $R_{\mbox{\tiny s}}+R_{\mbox{\tiny w}}-\hH_{\bx}(X) > 0$
is of the exponential order of $\exp\{-nE_{\mbox{\tiny sec}}(R_{\mbox{\tiny
s}}+R_{\mbox{\tiny w}})\}$,
where
\begin{equation}
E_{\mbox{\tiny sec}}(R)\dfn\min\{D(Q_X\|P_X):~H_Q(X)\le R\},
\end{equation}
and therefore,
\begin{eqnarray}
I(\bS;\bW)&\lexe&n\sum_{\bx}P_X(\bx)[R_{\mbox{\tiny s}}+R_{\mbox{\tiny
w}}-\hH_{\bx}(X)]\cdot\calI\{R_{\mbox{\tiny s}}+R_{\mbox{\tiny
w}}-\hH_{\bx}(X) > 0\}\nonumber\\
&\le&n(R_{\mbox{\tiny s}}+R_{\mbox{\tiny w}})\cdot\mbox{Pr}\{R_{\mbox{\tiny
s}}+R_{\mbox{\tiny w}} - \hH_{\bX}(X) > 0\}\nonumber\\
&\exe&\exp\{-nE_{\mbox{\tiny sec}}(R_{\mbox{\tiny s}}+R_{\mbox{\tiny
w}})\},
\end{eqnarray}
which means that as long as $H(X) > R_{\mbox{\tiny s}}+R_{\mbox{\tiny w}}$,
strong security is guaranteed in the sense that $I(\bS;\bW)$ tends to zero
even without normalization by $n$, as it decays exponentially fast. The secrecy exponent
depends on $R_{\mbox{\tiny s}}$ and $R_{\mbox{\tiny w}}$ only via their sum,
$R_{\mbox{\tiny s}}+R_{\mbox{\tiny w}}$.

\section*{Appendix}
\renewcommand{\theequation}{A.\arabic{equation}}
    \setcounter{equation}{0}

{\it Proof of eq.\ (\ref{appendixb}).}
The proof is similar to the proof of a similar argument in the context of
channel coding \cite[Appendix B]{p187}.
First, observe that
\begin{equation}
Z_{\bx}(\by)=\sum_{\bx'\ne
\bx}\exp\{na(\hat{P}_{\bx'\by})\}\cdot\calI\{f(\bx')=f(\bx)\}=\sum_{Q_{X|Y}}
e^{na(Q_{XY})}N(\calT(Q_{X|Y}|\by),f(\bx)).
\end{equation}
Thus, considering the randomness of $\{f(\bx)\}$,
\begin{eqnarray}
& &\mbox{Pr}\left\{Z_{\bx}(\by)\le
\exp\{n\alpha(R+\epsilon,\hat{P}_{\by})\}\right\}\nonumber\\
&=&\mbox{Pr}\left\{\sum_{Q_{X|Y}} N(\calT(Q_{X|Y}|\by),f(\bx))e^{na(Q_{XY})}\le
\exp\{n\alpha(R+\epsilon,\hat{P}_{\by})\}\right\}\nonumber\\
&\le&\mbox{Pr}\left\{\max_{Q_{X|Y}} N(\calT(Q_{X|Y}|\by),f(\bx))e^{na(Q_{XY})}\le
\exp\{n\alpha(R+\epsilon,\hat{P}_{\by})\}\right\}\nonumber\\
&=&\mbox{Pr}\bigcap_{Q_{X|Y}}\left\{N(\calT(Q_{X|Y}|\by),f(\bx))e^{na(Q_{XY})}\le
\exp\{n\alpha(R+\epsilon,\hat{P}_{\by})\}\right\}\nonumber\\
&=&\mbox{Pr}\bigcap_{Q_{X|Y}}\left\{N(\calT(Q_{X|Y}|\by),f(\bx))\le
\exp\{n[\alpha(R+\epsilon,\hat{P}_{\by})-a(Q_{XY})]\}\right\}.
\end{eqnarray}
Now, $N(\calT(Q_{X|Y}|\by),f(\bx))$ is a binomial random variable with 
$|\calT(Q_{X|Y}|\by)\exe e^{nH_Q(X|Y)}$
trials and
success rate of
$e^{-nR_{\mbox{\tiny w}}}$.
We now argue that
by the very definition of $\alpha(R+\epsilon,\hat{P}_{\by})$, there
must exist
some $Q_{X|Y}^*$ such that for $Q_{XY}^*=\hP_{\by}\times Q_{X|Y}^*$,
$H_{Q^*}(X|Y)\ge R+\epsilon$ and $H_{Q^*}(X|Y)-R-\epsilon \ge
\alpha(R+\epsilon,\hP_{\by})-a(\hP_{\by}\times Q_{X|Y}^*)$.
Let then $Q_{X|Y}^*$ be such a conditional distribution. Then,
\begin{eqnarray}
& &\mbox{Pr}\bigcap_Q\left\{N(\calT(Q_{X|Y}|\by),f(\bx))
\le \exp\{n[\alpha(R+\epsilon,\hP_{\by})-a(\hP_{\by}\times Q_{X|Y})]\}\right\}\nonumber\\
&\le&
\mbox{Pr}\left\{N(\calT(Q_{X|Y}^*|\by),f(\bx))\le
\exp\{n[\alpha(R+\epsilon,\hP_{\by})-a(\hP_{\by}\times Q_{X|Y}^*)]\}\right\}.
\end{eqnarray}
Now, we know that $H_{Q^*}(X|Y) \ge R+\epsilon$ and $H_{Q^*}(X|Y)-R-\epsilon
\ge
\alpha(R+\epsilon,\hP_{\by})-a(\hP_{\by}\times Q_{X|Y}^*)$.
By the Chernoff bound (see, e.g., \cite[Chap.\ 6]{fnt})ć,
the probability in question is upper bounded by
\begin{equation}
\exp\left\{-e^{nH_{Q^*}(X|Y)}D(e^{-\alpha n}\|e^{-\beta n})\right\},
\end{equation}
where $\alpha=H_{Q^*}(X|Y)+a(\hP_{\by}\times
Q_{XY}^*)-\alpha(R+\epsilon,\hP_{\by})$ and
$\beta=R$. Noting that $\alpha-\beta\ge\epsilon$,
we can easily lower bound the binary divergence as follows (see \cite[Section
6.3]{fnt}):
\begin{eqnarray}
D(e^{-\alpha n}\|e^{-\beta n})&\ge&e^{-\beta
n}\{1-e^{-(\alpha-\beta)n}[1+n(\alpha -\beta)]\}\nonumber\\
&\ge&e^{-nR}[1-e^{-n\epsilon}(1+n\epsilon)],
\end{eqnarray}
where in the last passage, we have used the decreasing monotonicity of the
function $f(t)=(1+t)e^{-t}$ for $t\ge 0$.
Thus,
\begin{eqnarray}
& &\mbox{Pr}\left\{N(\calT(Q_{X|Y}^*|\by),f(\bx))\le
\exp\{n[\alpha(R,\hP_{\by})-a(\hP_{\by}\times
Q_{X|Y}^*)-\epsilon]\}\right\}\nonumber\\
&\le&
\exp\left\{-e^{nH_{Q^*}(X|Y)}\cdot
e^{-nR}[1-e^{-n\epsilon}(1+n\epsilon)]\right\}\nonumber\\
&\le&\exp\left\{-e^{n\epsilon}[1-e^{-n\epsilon}(1+n\epsilon)]\right\}\nonumber\\
&=&\exp\left\{-e^{n\epsilon}+n\epsilon+1\right\}.
\end{eqnarray}
Finally, the factor of $|\calX\times\calY|^n$ in eq.\ (\ref{appendixb}) comes from the
union bound, taking into account all $|\calX\times\calY|^n$ possible pairs
$\{(\bx,\by)\}$. This completes the proof of eq.\ (\ref{appendixb}).

\end{document}